%% file: photons-pheno.tex
\documentclass[11pt]{article}

\usepackage{graphicx}
\usepackage{epsfig}
\usepackage{cite}
\usepackage{amssymb}
\usepackage{amsmath}
\usepackage{dsfont}
\usepackage{multirow}
\usepackage{ulem}
\usepackage{soul}
\usepackage{sidecap}
\include{preamble}

\usepackage{url} 
\usepackage{hyperref}
\usepackage{amsfonts}

\def\lsim{\mathrel{
   \rlap{\raise 0.511ex \hbox{$<$}}{\lower 0.511ex \hbox{$\sim$}}}}
\def\gsim{\mathrel{
   \rlap{\raise 0.511ex \hbox{$>$}}{\lower 0.511ex \hbox{$\sim$}}}}

\newcommand {\bq} {\begin{equation}}
\newcommand {\eq} {\end{equation}}

\def\smallfrac#1#2{\hbox{${{#1}\over {#2}}$}}
\newcommand{\be}{\begin{equation}}
\newcommand{\ee}{\end{equation}}
\newcommand{\bea}{\begin{eqnarray}}
\newcommand{\eea}{\end{eqnarray}}
\newcommand{\bi}{\begin{itemize}}
\newcommand{\ei}{\end{itemize}}
\newcommand{\ben}{\begin{enumerate}}
\newcommand{\een}{\end{enumerate}}

\newcommand{\lp}{\left(}
\newcommand{\rp}{\right)}

\def\frac#1#2{{{#1}\over {#2}}}
\def\gsim{\mathrel{\rlap{\lower4pt\hbox{\hskip1pt$\sim$}}
    \raise1pt\hbox{$>$}}}         %greater than or approx. symbol
\def\lsim{\mathrel{\rlap{\lower4pt\hbox{\hskip1pt$\sim$}}
    \raise1pt\hbox{$<$}}}         %less than or approx. symbol

\begin{document}
\begin{flushright}
IFUM-958-FT\\
Edinburgh 2010/17\\
\end{flushright}
\vspace{2.0cm}
\begin{center}
{\Large \bf
High energy resummation of direct photon
production  at hadronic colliders}
\vspace{0.8cm}

Giovanni~Diana$^{1}$, Juan~Rojo$^{1}$ and Richard~D.~Ball$^{2}$, 

\vspace{1.cm}
{\it  ~$^1$ Dipartimento di Fisica, Universit\`a di Milano and
INFN, Sezione di Milano,\\ Via Celoria 16, I-20133 Milano, Italy \\
~$^2$ School of Physics and Astronomy, University of Edinburgh,\\
JCMB, KB, Mayfield Rd, Edinburgh EH9 3JZ, Scotland}
%\end{center} 

\bigskip
\bigskip

%\begin{center}
{\bf \large Abstract:}
\end{center}

Direct photon production is an important process at hadron colliders, being
relevant both for precision measurement of the gluon density, and as 
background to Higgs and other new physics searches. Here we explore the 
implications of recently derived results for high energy resummation of 
direct photon production for the interpretation of measurements at 
the Tevatron and the LHC. The effects of
resummation are compared to various 
sources of theoretical uncertainties like
PDFs and scale variations. We show how the high--energy resummation procedure
 stabilizes the logarithmic enhancement of the cross 
section at high--energy which is present at any
fixed order in the perturbative expansion starting at NNLO. 
The effects of high--energy
resummation are found to be negligible at Tevatron, while they enhance
the cross section by a few percent for $p_T \lsim 10$ GeV at the LHC. 
Our results imply that the discrepancy at small $p_T$
between fixed order NLO and Tevatron data cannot be explained by
unresummed high--energy contributions.

\clearpage

The high--energy 
regime of QCD is the kinematical regime in
which hard scattering processes happen at a center-of-mass
energy $\sqrt{S}$ which is much larger than the characteristic hard scale
of the process $Q$. 
An understanding of strong interactions in this region
is therefore necessary in order to perform
precision physics at high--energy colliders. The high--energy regime
is also known as the small-$x$ regime, since it is the regime
in which the scaling variable $x=Q^2/S\ll 1$.
In this sense, HERA was
the first small $x$ machine, while at LHC the small $x$ regime will 
be even more important. 

As is well known, deep--inelastic partonic cross sections and parton
splitting functions
receive large corrections in the small $x$ limit due to the presence of powers 
of $\alpha_s\log x$ to all orders in the perturbative 
expansion~\cite{Catani:1990eg,Catani:1994sq}.
This suggests dramatic effects from yet higher
orders, so the success of NLO perturbation theory 
at HERA was
for a long time very hard to explain.
In the last several years this situation 
has been clarified~\cite{Altarelli:2003hk,Altarelli:2005ni,Ball:2007ra,Altarelli:2008aj,Ciafaloni:2003rd,Ciafaloni:2003kd,Ciafaloni:2007gf},
showing that, once the full resummation procedure
accounts for running coupling effects, gluon exchange symmetry and
other physical constraints, the effect of
the resummation of terms which are enhanced
 at small $x$ is
perceptible but moderate --- comparable in size to typical NNLO fixed
order corrections in the HERA region. 

A major development for high--energy
resummation was presented in Ref.~\cite{Altarelli:2008aj} 
where the full small $x$ resummation
of deep-inelastic scattering (DIS) anomalous dimensions and
coefficient functions was obtained including quarks, 
which allowed for the first time a consistent
small-$x$ resummation of DIS structure functions.
Furthermore, the resummation of hard partonic cross sections
has been performed for several LHC processes such as 
heavy quark production~\cite{Ball:2001pq}, 
Higgs production~\cite{Marzani:2008az,Marzani:2008ih},
Drell-Yan~\cite{Marzani:2008uh,Marzani:2009hu} 
and prompt photon production~\cite{Diana:2009xv}. 
Hints of the presence of small-$x$ resummation 
have also recently found in inclusive HERA data~\cite{Caola:2009iy}.
Small--$x$ resummation should also be very important
at a high--energy DIS collider like the Large
Hadron Electron Collider~\cite{Rojo:2009us,Rojo:2009ut}.
A more detailed summary of
recent theoretical developements in high--energy resummation may be found in 
Ref.\cite{Forte:2009wh}.
These results mean that a detailed analysis of the
impact of high--energy resummation on precision LHC physics
is now possible. 

As a part of such a program, in this letter we present a study of the
 phenomenological implications
of the high--energy resummation of direct photon production at hadronic
colliders. The production of direct photons~\cite{Owens:1986mp} is
a very important process at hadronic colliders, relevant
both for fundamentals reasons (tests of perturbative QCD, measurement of the gluon PDF)
and as background to new physics searches, the $H\to \gamma \gamma$
decay being the classical example.
In the case of direct photon production, several works have studied in 
detail the comparison
of theoretical QCD predictions with available experimental data from
fixed target and collider experiments. Such
comparisons have been performed using fixed
order NLO computations~\cite{Huston:1995vb,Aurenche:2006vj,ddephotons},
Monte Carlo event generators~\cite{Hoeche:2009xc} and supplementing
the fixed order result with  threshold
resummations~\cite{Catani:1999hs,Kidonakis:1999hq,Bolzoni:2005xn,
Becher:2009th}. The
latter 
aim to improve the accuracy of the perturbative
prediction in the regime where the photon's 
$p_T$ is large, close to the kinematic
production threshold, where soft gluon emission enhances the cross
section. 

The present work is instead focused on the
low $p_T$ region, where terms of the type $\alpha_s^k\ln^p x$,
enhanced by logarithms of the scaling variable 
$x_{\perp}\equiv 4p_T^2/S$, are 
important to all orders in perturbation theory.
 For this reason we do not
consider fixed target data, which are characterized by moderate and
large values of $x_{\perp}$ where high--energy resummation is 
certainly irrelevant,
and concentrate instead on collider data for which the large 
center of mass $S\gg p_T^2$ available guarantees that
the kinematical region sensitive to small-$x$ effects is explored.
As an illustration, if the 
small-$p_T$ region is defined naively as the region
in which the hadronic cross section becomes sensitive to PDFs and
partonic coefficient
functions for $x\lsim 10^{-3}$, then at Tevatron this criterion corresponds
to $p_T \lsim 30$~GeV and at the LHC~14~TeV to $p_T \lsim 200$~GeV.

\begin{figure}[t!]
\begin{center}
\epsfig{width=.44\textwidth,figure=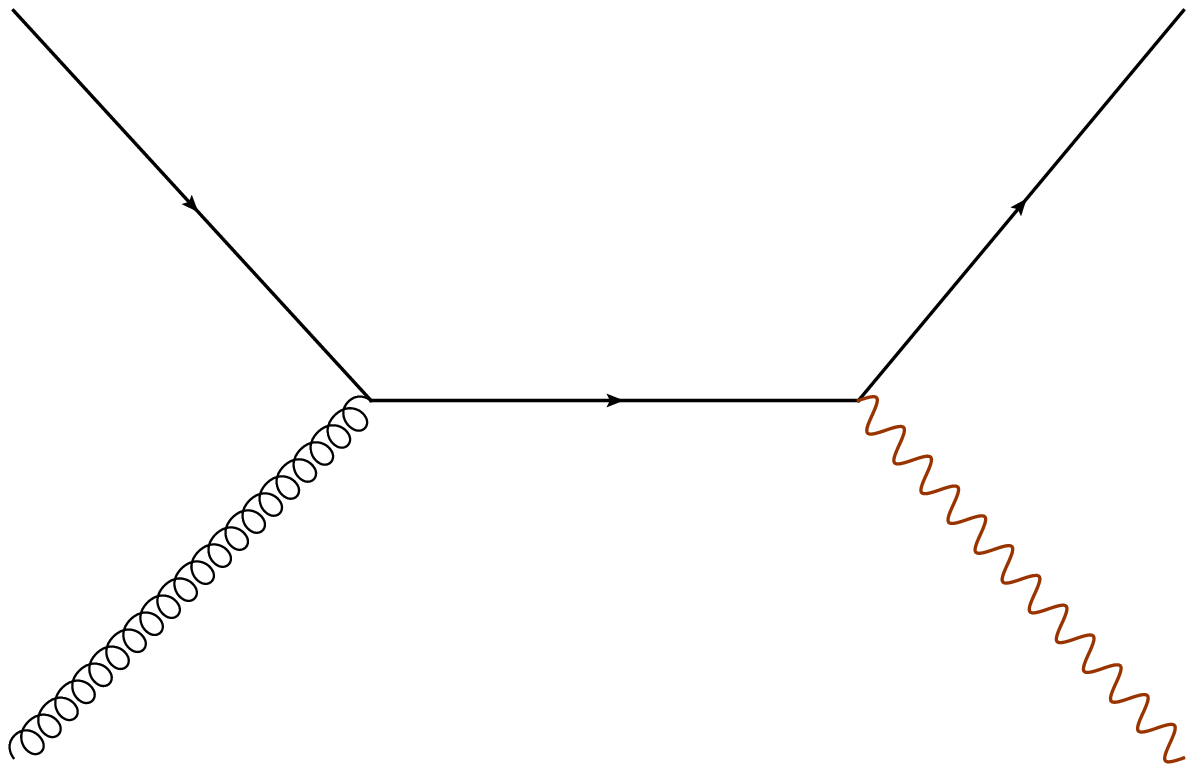} $\qquad$
\epsfig{width=.44\textwidth,figure=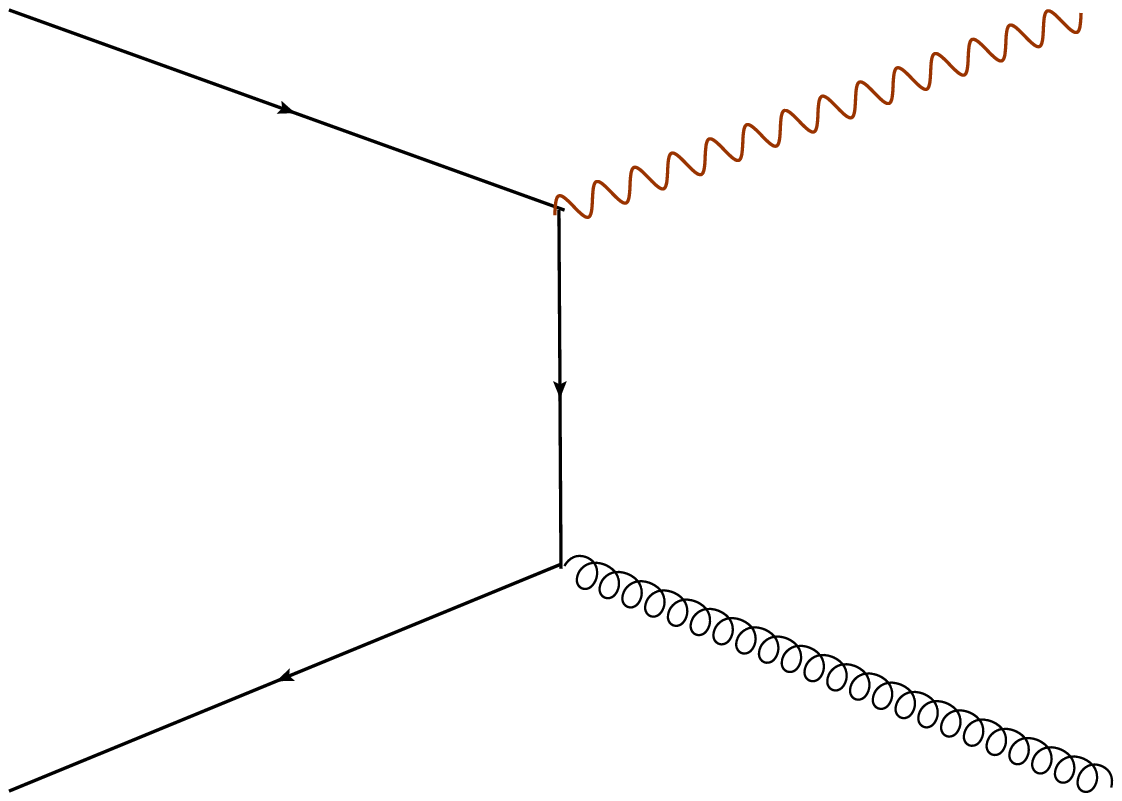} 
\caption{\small The Feynman diagrams for the
direct production of a photon in hadronic collisions
at leading order: the $gq$ channel, also known as
Compton scattering channel (left) and the $q\bar{q}$
channel, also known as quark annihilation channel (right). 
\label{fig:feyn} }
\end{center}
\end{figure}

The prompt photon process is characterized by a hard event involving the production of a single photon. Let us consider the hadronic process
\begin{equation}
H_1(P_1)+H_2(P_2)\rightarrow \gamma(q)+X.
\end{equation}
According to perturbative QCD, the direct and the fragmentation component 
of the inclusive cross-section at fixed transverse momentum $p_T$ 
of the photon 
can be written as~\cite{Catani:1999hs}
\begin{eqnarray} 
&&p_T^3 \frac{d\sigma_\gamma(x_\perp,p_T^2)}{d p_T}=\sum_{a,b}\int_{x_\perp}^1dx_1\; f_{a/H_1}(x_1,\mu^2_{\rm F})\int_{x_\perp/x_1}^1 dx_2\;f_{b/H_2}(x_2,\mu^2_{\rm F})\nonumber\times\\&&\times\int_0^1 dx\left\lbrace\delta\left(x-\frac{x_\perp}{x_1 x_2}\right)  \mathcal{C}^\gamma_{ab}(x,\alpha_s(\mu^2);p_T^2,\mu_{\rm F}^2,\mu_{\rm f}^2)+\nonumber\right.\\
&&\left.+\sum_c\int_0^1 dz\;z^2d_{c/\gamma}(z,\mu^2_f)\delta\left(x-\frac{x_{\perp}}{zx_1x_2}\right)\mathcal{C}^c_{ab}(x,\alpha_s(\mu^2);p_T^2,\mu_F^2,\mu_f^2)\right\rbrace, \label{eq:fact}
\end{eqnarray}
where we have introduced the customary scaling variable in terms of the hadronic center-of-mass energy $S=(P_1+P_2)^2$ :
\begin{equation}
x_\perp=\frac{4 p_T^2}{S},  \qquad 0<x_\perp<1 \ .
\end{equation}
The fragmentation component is 
given in terms of a convolution with the 
fragmentation function $d_{c/\gamma}(z,\mu^2_f)$.
In the factorization formula Eq.~(\ref{eq:fact}) we have used the short-distance cross-sections
\begin{eqnarray}
\mathcal{C}^{\gamma(c)}_{ab}&\equiv &p_T^3\frac{d\hat\sigma_{ab\rightarrow\gamma(c)}(x,\alpha_s(\mu^2);p_T^2,\mu_F^2,\mu_f^2)}{dp_T},\label{eq:coefdef}
\end{eqnarray}
where $a$, $b$ and $c$ are parton indices ($q$, $\bar q$, $g$)
 while $f_{i/H_j}(x_i,\mu^2_F)$ is the parton density at the 
factorization scale $\mu_F$. The leading
order coefficient functions for the Compton
scattering channel ($qg$) and for  the quark annihilation channel
($q\bar{q}$) are given by
\bea
\mathcal{C}^{\gamma,{\rm LO}}_{qg}(x)&= &\frac{\alpha\alpha_se_q^2\pi}{2N_c}
\frac{x}{\sqrt{1-x}}\lp 1+\frac{x}{4}\rp \ ,\nonumber \\
\mathcal{C}^{\gamma,{\rm LO}}_{q\bar{q}}(x)&= &\frac{\alpha\alpha_se_q^2C_F\pi}{N_c}
\frac{x}{\sqrt{1-x}}\lp 2-x\rp \ .
\eea
In Fig.~\ref{fig:feyn} we show the associated LO Feynman
diagrams for these two channels. NLO corrections
to the direct partonic cross section in Eq.~(\ref{eq:fact}) were computed
in Refs.~\cite{NLOdir1,NLOdir2,Gordon:1993qc},
while for the fragmentation component they were evaluated in 
Refs.~\cite{NLOfrag1,NLOfrag2}.

%------------------------------------------------------------
%------------------------------------------------------------------

\begin{figure}[t!]
\begin{center}
\epsfig{width=0.85\textwidth,figure=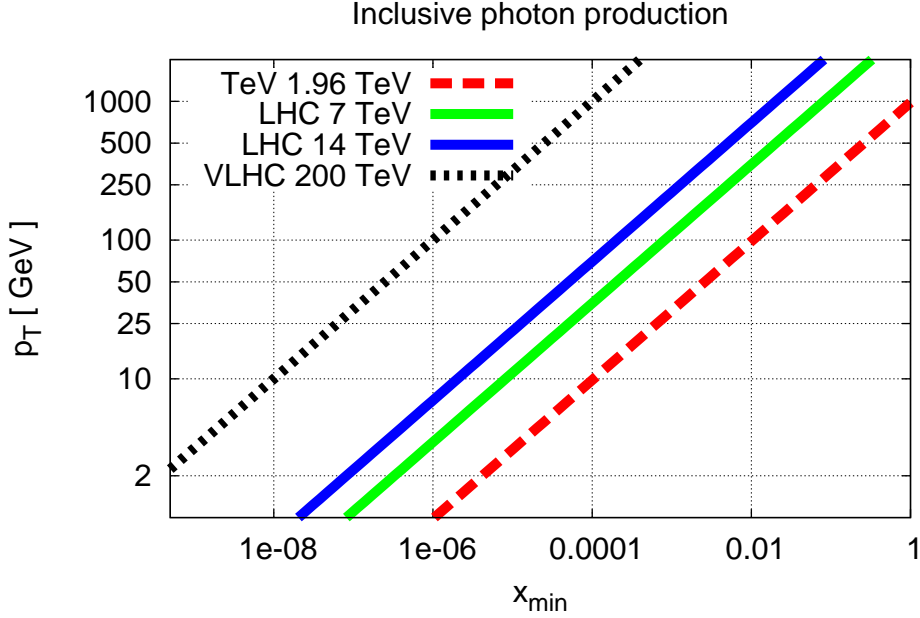}
\caption{\small The minimum values of $x$, $x_{\rm min} =x_{\perp}=4p_T^2/S$
which are probed in the production of a direct photon with
transverse momentum $p_T$ at hadronic colliders: Tevatron
Run II ($\sqrt{S}$=1.96 TeV), LHC 7 TeV and LHC 14 TeV and
VLHC 200 TeV. 
As can be seen from the plot, for the production of a $p_T\sim 20$ GeV photon,
 PDFs and coefficient functions are probed down to 
$x\sim 5\,10^{-4}$ at the Tevatron and
$x\sim 10^{-5}$ at the
LHC 14 TeV. Note that no cuts in rapidity are assumed in the
definition of the kinematical ranges, experimentally realistic cuts
reduce the reach in $x$ for a given $p_T$.}
\label{fig:kin}
\end{center}
\end{figure}

The kinematics of direct photon production at hadronic
colliders are summarized in Fig.~\ref{fig:kin}, where the
minimum value of $x$, $x_{\perp}$, probed in the
production of a photon with a given $p_T$ is shown.
For illustrative purposes, the corresponding kinematics
for a notional VLHC with $\sqrt{S}=200$ TeV are also shown.
 From
 Fig.~\ref{fig:kin} follows that collider experiments
have the potential reach down to very small values of $x$,
for example, at
LHC 14 TeV  PDFs and coefficient functions are probed down
to $x\sim 10^{-5}$ for a $p_T\sim 20$ GeV photon. This implies that
one should worry about those terms in the perturbative expansion
which are formally subleading but which are logarithmically enhanced
to all orders at small-$x$, both in the PDF evolution and
in the partonic cross-sections.

%------------------------------------------------------------
%------------------------------------------------------------------

Due to multiple gluon emissions, the perturbative expansion of the 
partonic cross sections, Eq.~(\ref{eq:coefdef}), 
is logarithmically enhanced at small-$x$ starting from
NNLO. While at
 NLO the single gluon emission produces the constant behaviour at 
low-$x$ of the coefficient function Eq.~(\ref{eq:coefdef}), the NNLO
 behaves like a single logarithm and, in general, at N$^k$LO, the 
dominant contribution is given by $\alpha_s(\alpha_s \log x)^{k-1}$.

The high--energy resummed coefficient function of the direct component 
in Eq.~(\ref{eq:fact}) has been obtained in Ref.~\cite{Diana:2009xv} 
in the framework of the $k_T$-factorization theorem, which allows one to 
 perform the leading log resummation in terms of the off shell 
impact factor, which is the leading order cross section computed with 
off-shell incoming gluons.  Following the resummation procedure 
one obtains the sum of the leading contributions at high--energy 
and, by re-expanding in powers of $\alpha_s$, we have the 
coefficients of each power of $\log x$  to all orders in
 perturbation theory.

The high-energy enhanced terms in the direct photon partonic
cross section, as discussed in  Ref.~\cite{Diana:2009xv},
in $N$ space are given in the $qg$ channel by
\be
\tilde{\mathcal{C}}^\gamma_{qg}(N,\bar{\alpha}_s,\kappa_r)= 
\frac{\alpha\alpha_s^2}{N}\sum_{k=0}^{\infty} c_{qg}^{(k)}\lp \kappa_r \rp\lp\frac{\bar{\alpha}_s}{N}\rp^{k-1} \label{eq:res2}
\ee
where the renormalization scale has been set
to proportional to the transverse momentum
of the photon $\mu_{\rm r}= \kappa_r p_T$ 
%$\mu$ is always dimensionful
and
where $\bar{\alpha}_s\equiv \alpha_sC_A/\pi$ with $\alpha_s$ 
is the fixed strong coupling and
 $\alpha=1/137$ the electromagnetic coupling constant. The first few
coefficients in Eq.~(\ref{eq:res2}) read
\begin{eqnarray}
c_{qg}^{(0)}&=&\smallfrac{7}{6} \nonumber \\
c_{qg}^{(1)}&=&\smallfrac{67}{36}-\smallfrac{7}{3}\log
   \kappa_r \nonumber \\
c_{qg}^{(2)}&=&\smallfrac{7}{4} \log ^2\kappa_r -\smallfrac{29}{9} \log
      \kappa_r +\smallfrac{385}{216}\nonumber\\
c_{qg}^{(3)}&=& -\smallfrac{7}{9}\ln^3\kappa_r-\smallfrac{55}{26}\ln^2\kappa_r
-\smallfrac{179}{54}\ln \kappa_r
+\smallfrac{49}{9} \zeta(3)+\smallfrac{2323}{1296}  \label{eq:rescoeff}
%c_{qg}^{(4)}&=&\frac{(6 \log \kappa_r (3 \log \kappa_r (\log \kappa_r
%               (105 \log \kappa_r-428)+1047)-11088 \zeta (3)-4484)+59760 \zeta
%                  (3)+7429)}{7776}\nonumber
\end{eqnarray}
The NLO term in Eqs.~(\ref{eq:res2}-\ref{eq:rescoeff}) 
gives, in the $x$-space, 
the constant value 
$\alpha{\alpha}_s^267/36$ for $\kappa_r=1$,
in agreement with the fixed order calculation of 
Refs.~\cite{Gordon:1993qc,Ellis:1990hw}.
By using the high-energy color charge relation between the hard coefficient 
functions
\be
\tilde{\mathcal{C}}^\gamma_{q\bar q(q)}(N,\alpha_s,\kappa_r)
=\smallfrac{C_F}{C_A}\big(\tilde{\mathcal{C}}^\gamma_{qg}(N,\alpha_s,\kappa_r)
-\tilde{\mathcal{C}}^{\gamma,{\rm LO}}_{qg}(0,\alpha_s,\kappa_r)\big)
\ee
we can obtain the high--energy coefficient function in the 
$q\bar{q}(q)$ channel.

In the rest of this work we will set $\kappa_r=1$. In this case, the
resummed coefficient function Eq.~(\ref{eq:res2}) in $x-$space  reads
\begin{eqnarray}
\mathcal{C}^\gamma_{qg}(x,\bar{\alpha}_s)&=&\alpha\alpha_s^2
\big\{\smallfrac{67}{36} +\smallfrac{385
   }{216} \bar{\alpha}_s\ln\smallfrac{1}{x}
+ \smallfrac{1}{2}(\smallfrac{2323}{1296}+\smallfrac{49}{9}\zeta(3))
\bar{\alpha}_s^2\ln^2\smallfrac{1}{x}
\nonumber \\
   &+&\smallfrac{1}{6}(\smallfrac{14233}{7776}-\smallfrac{7}{720}\pi^3
+\smallfrac{308}{27}\zeta (3))
   \bar{\alpha}_s^3\ln^3\smallfrac{1}{x} 
+ O(\bar{\alpha}_s^4\ln^4\smallfrac{1}{x})\big\}.\label{eq:res1}
\label{eq:coeffres}
\end{eqnarray}
Note that the logarithms of $x$ (high--energy enhanced terms) which lead
to the rise of the partonic cross section at small-$x$ appear
only from NNLO onwards.

However, this formalism is incomplete because it does not account for
running coupling effects. Indeed, as shown in Refs.~\cite{Ciafaloni:2003rd,Ciafaloni:2007gf,Altarelli:2005ni,Ball:2007ra}
the running of $\alpha_s$ produces a new series of relevant contributions
in the high energy limit which modify the nature of the 
singularity of the anomalous dimension at small-$x$. 
At fixed $\alpha_s$, the resummation procedure requires the identification 
of the Mellin variable $M$ (conjugate of $Q^2$) with the sum 
of the leading singularities of the resummed anomalous dimension
\be
M=\gamma_s\left(\alpha_s/N\right).\label{eq:ident}
\ee
Now, if we include running effects, $\alpha_s$ becomes a function of $Q^2$
which corresponds to an operator in $M$-Mellin space and Eq.~(\ref{eq:ident})
is understood as an equality between operators. At the running coupling level, 
the identification given by Eq.~(\ref{eq:ident}) produces a class of terms 
proportional to increasing derivatives of $\gamma_s$. In practice these are 
most easily computed by using Eq.~(\ref{eq:ident}) to turn the expansion 
Eq.~(\ref{eq:res2}) 
in powers of $\bar\alpha_s/N$ into an expansion in powers of 
$M$: since powers of $m$ correspond to derivatives with respect to 
$\ln Q^2$, this then gives the resummed coefficient function even when the 
coupling runs. For a thorough description of the inclusion 
of running coupling effects see Refs.~\cite{Ball:2007ra,Altarelli:2008aj}.

In this way a fully resummed coefficient 
function in the $\overline{\rm MS}$ scheme, which can be consistently
matched to standard $\bar{\rm MS}$ fixed order computations, can be
obtained. This resummed coefficient function 
$\mathcal{C}^{\gamma,{\rm res}}_{ab}$ can be matched to  the
fixed order NLO coefficient function to obtain
a resummed coefficient functions which reproduces at large-$x$
the fixed order result,
\be
\mathcal{C}^{\gamma,{\rm NLOres}}_{ab}=
\mathcal{C}^{\gamma,{\rm NLO}}_{ab} + \mathcal{C}^{\gamma,{\rm res}}_{ab}-
\mathcal{C}^{\gamma,{\rm dc}}_{ab} \ .
\label{eq:cres}
\ee
In Eq.~(\ref{eq:cres}) the matching between  the fixed order NLO
result and the resummed one has been performed being careful of
 avoiding double counting. Therefore, the double counting
contribution $\mathcal{C}^{\gamma,{\rm dc}}_{ab}$, that is,
the terms in Eq.~(\ref{eq:coeffres}) up to $\mathcal{O}\lp \alpha\alpha_s^2\rp$
 is removed from
the NLO coefficient functions. The fixed order NLO
coefficient functions are taken from Ref.~\cite{Gordon:1993qc}. 

Note that Eq.~(\ref{eq:cres}) accounts for the high--energy resummation
of the direct part of the photon production cross section
without photon isolation effects.  At the resummed level, the effects of 
the photon isolation in the coefficient function Eq.~(\ref{eq:res2})
 can be computed 
in the small cone approximation. It can be shown that isolation
leads 
 an effect analogous to the variation of the renormalization scale, that
is, using the usual isolation with a cone of radius $R$ implies a
modifications of the renormalization scale
$\kappa_r \to \kappa_r R$. Note also that we
 do not attempt a resummation of the fragmentation
component of the photon production cross section, which in any case
is very much suppressed by the photon isolation. 
%The resummation
%of the fragmentation part is technically much more challenging and
%it is equivalent to the resummation of jet production in hadronic
%collisions~\cite{zaro}.

\begin{figure}[t!]
\begin{center}
\epsfig{width=.82\textwidth,figure=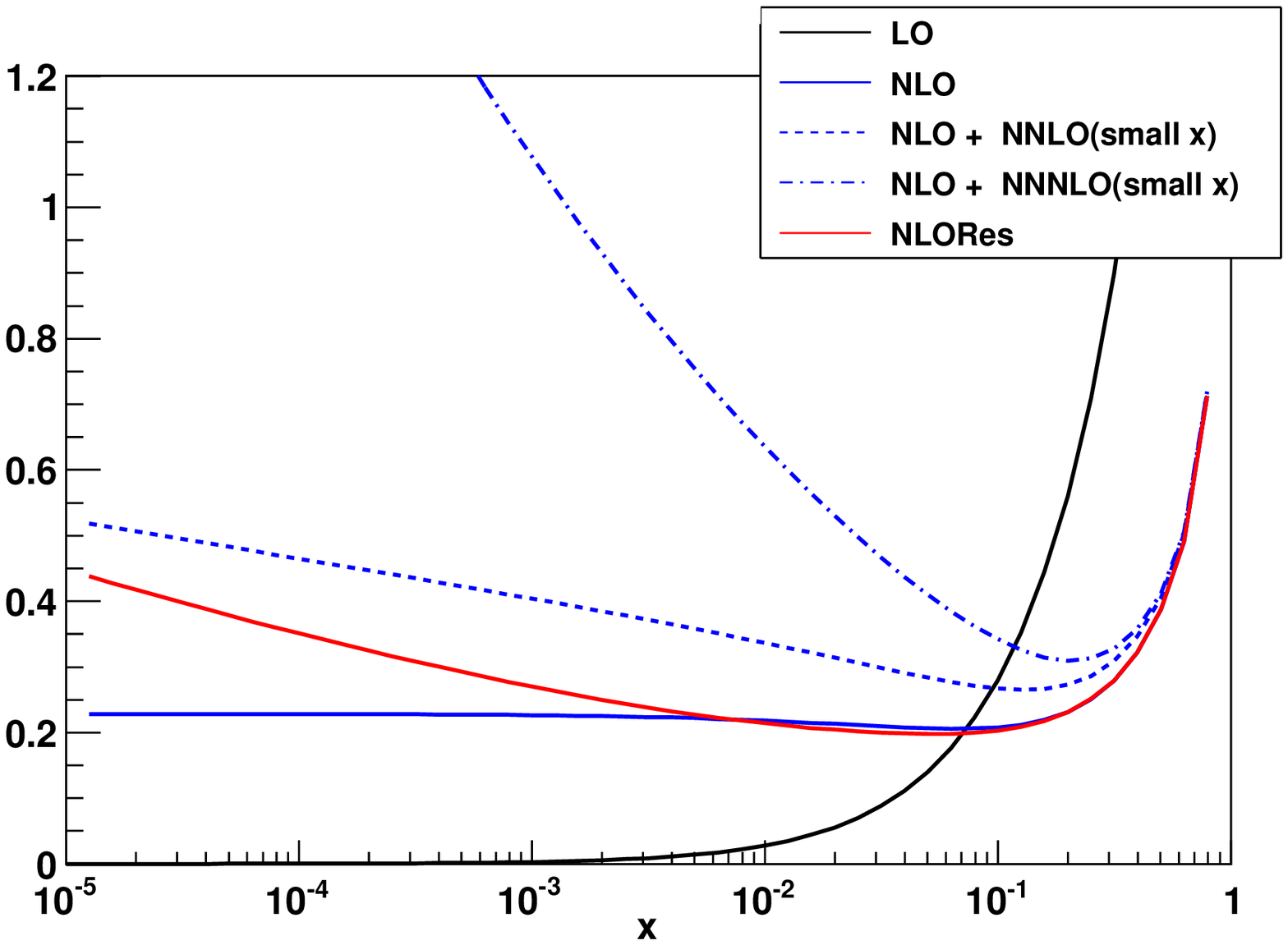}
\epsfig{width=.82\textwidth,figure=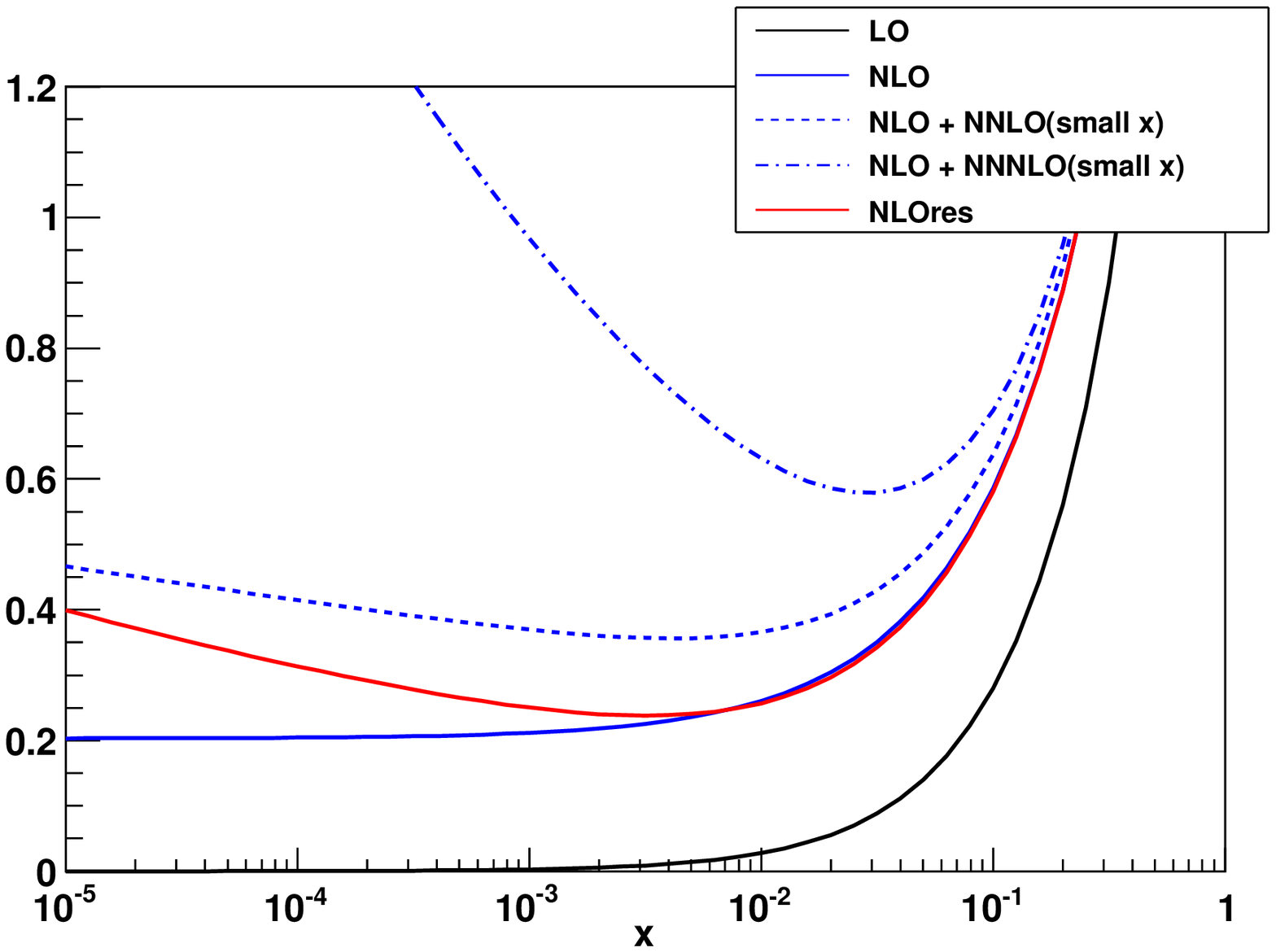} 
\caption{\small Upper plot: the coefficient functions 
(partonic cross sections) for direct photon production in
the $qg$ (Compton) channel. The following approximations to 
the partonic cross section are
shown: LO (black, solid), NLO (blue, solid), NLO with the addition
of the dominant  small-$x$  NNLO terms (blue, dashed), NLO with the addition
of the dominant  small-$x$  NNNLO terms (blue, dot-dashed), and finally
the high--energy resummed coefficient function, suitably
matched of the fixed order NLO Eq.~(\ref{eq:cres}) (red, solid).
Lower plot: the same comparison for the coefficient functions
in the $q\bar q$ (quark annihilation) channel. Note that the
coefficient functions rise at small-$x$ begins at NNLO only.\label{fig:mycoef}}
\end{center}
\end{figure}

In Fig.~\ref{fig:mycoef} we show the 
LO, NLO and resummed coefficient 
functions for the two relevant 
channels: Compton scattering, $qg$, and quark annihilation, $q\bar q$. 
On top of these, we also
show the NLO coefficient functions supplemented by the
NNLO high--energy contribution (the $\mathcal{O}\lp \alpha\alpha_s^3\rp$ term
in Eq.~(\ref{eq:coeffres}), and similarly 
for NLO plus NNNLO high--energy contributions  
(the $\mathcal{O}\lp \alpha\alpha_3^3\rp$ and $\mathcal{O}\lp \alpha\alpha_s^4\rp$ terms
in Eq.~(\ref{eq:coeffres}). These two latter cases are shown for
illustration, with the well know caveat that subleading corrections
at a fixed $\alpha_s$
might sizably reduce the effect of the leading high--energy contributions.
Fig.~\ref{fig:mycoef} shows the
important result that the steep rise
at small-$x$ of the fixed order coefficient function due to the
increasing powers of $\log x$ is 
stabilized after the including the running coupling effects,
as happens for DIS~\cite{Altarelli:2008aj}.

%------------------------------------------------------------
%------------------------------------------------------------------

Now that the resummed partonic cross-section, suitably matched
to the fixed-order NLO result, has been obtained, we can use
it to estimate the impact of high--energy resummation on the hadronic
cross--section, Eq.~(\ref{eq:fact}), at the Tevatron and at the
LHC.
The fixed order NLO computation of isolated
photon production  has been obtained  
using the code of Ref.~\cite{Gordon:1993qc}.
The small cone approximation for the isolation criterion
has been used, which is shown to be an excellent 
approximation~\cite{Catani:2002ny}
to the exact result for typical isolation parameters.
The photon fragmentation functions are the BFG set~\cite{PhotFrag},
although the choice is irrelevant since the fragmentation
component is severely suppressed by the isolation criterion.

Note that in the following the same PDF set will be used used both in the
NLO and in the resummed computations. The motivation for this is that 
we are interested only in the impact of the resummation of the partonic
cross-section. A consistent high--energy resummed cross section
would require PDFs obtained from a global analysis based on small-$x$
resummation, which are not available yet.

\def\GeV{\rm GeV}
In order to assess the impact of high--energy resummation
at the Tevatron, 
we consider  recent Run II data on isolated photon production 
from the CDF collaboration~\cite{Aaltonen:2009ty}. CDF data
is provided in the range $30~\GeV \le p_T \le 350~\GeV$, integrated
in the photon's rapidity range $|\eta^{\gamma}|\le 1.0$. 
The parameters of the photon
isolation criterion in the theoretical calculation match those
of the experimental analysis, namely $R=0.4$ and $E_{T}^{\rm had}\le 2~\GeV$.
The parton distribution set used for the comparison with
experimental data  is the
recent NNPDF2.0 global analysis~\cite{Ball:2010de}.
As compared
to previous NNPDF sets~\cite{DelDebbio:2007ee,Ball:2008by,Ball:2009mk,Ball:2009qv},
 NNPDF2.0  has a more precise gluon
both at small-$x$ from the combined HERA-I dataset and at large-$x$ 
from the Tevatron inclusive jet data, which translate
into very accurate predictions for direct photon
production.

\begin{figure}[t!]
\begin{center}
\epsfig{width=.8\textwidth,figure=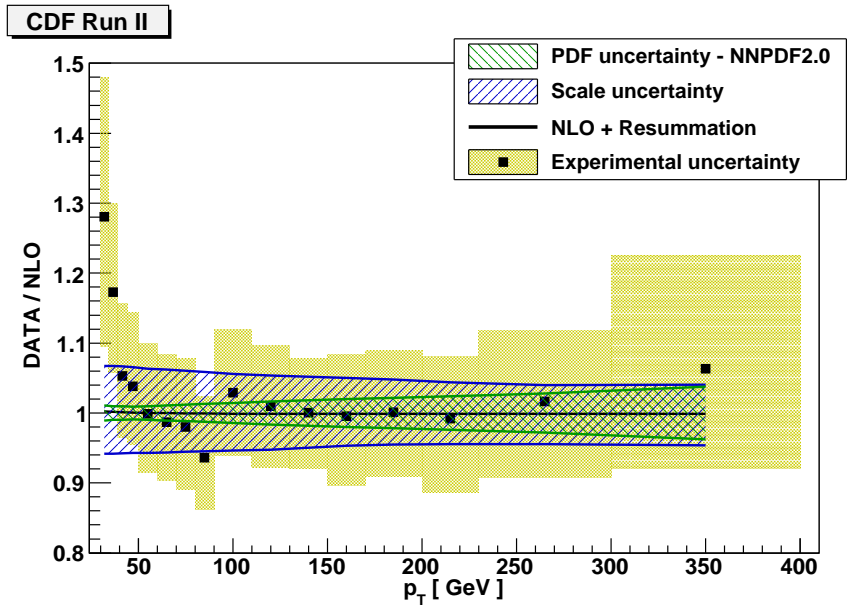} 
\caption{\small Comparison between the NLO 
cross section and the recent CDF data using the 
NNPDF2.0 PDF set. The solid black line is the
ratio between the high--energy resummed result and the NLO prediction,
as can be seen, the two results are essentially identical. The scale
variation uncertainty corresponds to the NLO calculation. }
\label{fig:NLOpheno}
\end{center}
\end{figure}

In Fig.~\ref{fig:NLOpheno} we present the results of
this comparison between the fixed order NLO and
the resummed predictions with the recent direct photon
measurements from the CDF Collaboration at Run II.
We show as well the PDF uncertainties and the theoretical uncertainties
from missing higher orders estimated as usual varying the scales
of the NLO expressions.
Good agreement between NLO QCD and experimental data within
the experimental uncertainties is found through most
of the $p_T$ range, except for a systematic discrepancy at small
$p_T$. This discrepancy is present also for other PDF
sets~\cite{ddephotons} as well as for the D0 data~\cite{Abazov:2005wc}. 

Since the high--energy resummed coefficient functions, 
Fig.~\ref{fig:mycoef} are
integrated in the photon's rapidity $\eta^{\gamma}$, we will assume that the effects of the
resummation are constant in $\eta^{\gamma}$. This means that the
resummed result in Fig.~\ref{fig:NLOpheno} has been obtained
as follows
\be
\frac{d\sigma^{\rm res}_{\gamma}\lp x_{\perp},p_T^2,
|\eta^{\gamma}|\le \eta_{\rm cut}\rp}{dp_T} 
= \frac{d\sigma^{\rm NLO}_{\gamma}\lp x_{\perp},p_T^2,
|\eta^{\gamma}|\le \eta_{\rm cut}\rp}{dp_T}
\frac{d\sigma^{\rm res}_{\gamma}\lp x_{\perp},p_T^2\rp}
{d\sigma^{\rm NLO}_{\gamma}\lp x_{\perp},p_T^2\rp}
\ee
This approximation could be improved by computing the high--energy resummation
of the photon rapidity distribution, for which qualitative arguments suggest
that the impact of resummation is more important towards forward
rapidities.

To estimate the theoretical uncertainty due to missing higher orders terms
in the NLO computation 
the  common scale $\kappa_r=\kappa_F=\kappa_f$
has been varied within a reasonable range. In particular we have
computed the cross section for $\kappa_r=$ 0.5, 1 and 2.
The scale variation uncertainty is defined as the envelope of the
most extreme results obtained this way for any given $p_T$.
As seen in  Fig.~\ref{fig:NLOpheno}, PDF uncertainties for isolated photon
production at the Tevatron are below 5\% in all the
$p_T$ range, and $\mathcal{O}\lp 2\%\rp$ in the small
$p_T\lsim 100$ GeV region. Scale variation uncertainties are
 $\mathcal{O}\lp 5\%\rp$ approximately constant in $p_T$.

We do not attempt here to estimate the combined 
PDF and $\alpha_s$ 
uncertainty~\cite{Demartin:2010er,Martin:2009bu,Lai:2010nw}, which 
could be important
in direct photon production since the cross section starts 
at $\mathcal{O}\lp \alpha\alpha_s\rp$.
Moreover, in this work we do not address the important
issue of the compatibility of predictions obtained
from different modern PDF sets, which has
already been presented in detail in Ref.~\cite{ddephotons}.

From Fig.~\ref{fig:NLOpheno} it is clear that at the Tevatron the
prediction from high--energy resummation is essentially identical
to that of the fixed order NLO computation. This might seem unintuitive,
since we have shown in Fig.~\ref{fig:mycoef} that the
respective coefficient functions are rather different 
in the small $x$ region within
the kinematical reach of experimental data (Fig.~\ref{fig:kin}). 
In order to explain this result,
let us define
the contribution to the total cross section 
for $x \ge x_{\perp}^{\rm min}$ as follows
\begin{eqnarray} 
&&\q^3 \frac{d\sigma_\gamma(x_\perp,x_\perp^{\rm min},p_T^2)}{d\q}\equiv \sum_{a,b}\int_{x_\perp^{\rm min}}^1dx_1\; f_{a/H_1}(x_1,\mu^2_{\rm F})\int_{x_\perp^{\rm min}/x_1}^1 dx_2\;f_{b/H_2}(x_2,\mu^2_{\rm F})\nonumber\times\\&&\times\int_0^1 dx\left\lbrace\delta\left(x-\frac{x_\perp}{x_1 x_2}\right)  \mathcal{C}^\gamma_{ab}(x,\alpha_s(\mu^2);p_T^2,\mu_{\rm F}^2,\mu_{\rm f}^2)+{\rm fragmentation}
\right\rbrace
\label{eq:ratio}
\end{eqnarray}
and then we can construct the ratio
\be
R_{\gamma}\lp x_{\perp}, x_{\perp}^{\rm min}, p_T^2 \rp
\equiv \frac{ d\sigma_\gamma(x_\perp,x_\perp^{\rm min},p_T^2)/dp_T}{
d\sigma_\gamma(x_\perp,x_\perp,p_T^2)/dp_T}
\ee
which measures the fraction of the cross--section for which
PDFs and coefficient functions with $x\ge  x_{\perp}^{\rm min} $
are probed. 

\begin{figure}[t!]
\begin{center}
\epsfig{width=.8\textwidth,figure=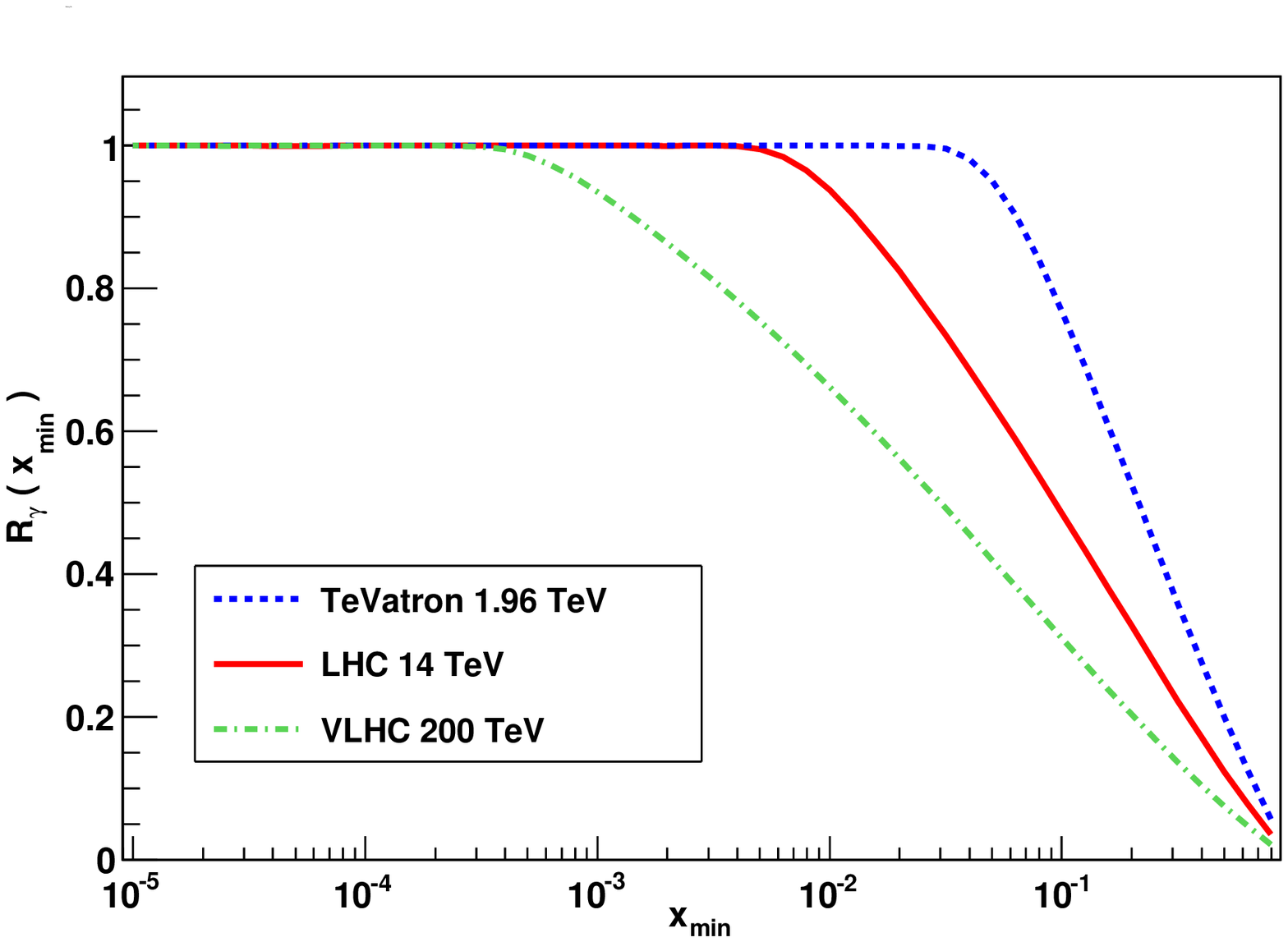} 
\caption{\small The ratio $R_{\gamma}$, Eq.~(\ref{eq:ratio}), as a function
 of $x_{\perp}^{\rm min}$
at the LHC $\sqrt{S}=14$ TeV (red solid line) and
at Tevatron Run II $\sqrt{S}=1.96$ TeV  (blue dashed line)
for the production of photon with $p_T=20$ GeV. It is clear that
the cross-section is dominated by the contribution of the
coefficient function at medium and large-$x$, 
$x\gsim 5\,10^{-3}$ for LHC and $x\gsim 5\,10^{-2}$
for the Tevatron. The fact that the total cross section
is insensitive to the partonic cross-sections at small-$x$ explains
the reduced impact of the high--energy resummation at hadronic
colliders.}
\label{fig:ratio_xsec_lhc}
\end{center}
\end{figure}

In Fig.~\ref{fig:ratio_xsec_lhc} we show this ratio 
at the Tevatron, the LHC and the notional VLHC
  for the production of a photon 
with $p_T=20$ GeV. We observe that
the direct photon cross section at the Tevatron  is
completely dominated by the region $x\gsim 5\,10^{-2}$. In this
region, the resummed coefficient functions are almost
identical to the fixed order NLO ones.
Therefore, despite the fact that the values of $x$ probed
in small-$p_T$ photon production are such
that the resummed coefficient functions, 
Fig.~\ref{fig:mycoef},
differ sizably from their fixed order NLO counterparts, this
difference is restricted to a region with very little weight in
the total cross--section. This feature
of direct photon production 
(shared also by Higgs production~\cite{Marzani:2008az,Marzani:2008ih})
explains the smallness
of high--energy resummation at the Tevatron. Note that
this applies to the rapidity integrated cross--section, it is
conceivable that more important effects are observed if
one is restricted to forwards rapidities.

%------------------------------------------------------------
%------------------------------------------------------------------

%------------------------------------------------------------
%------------------------------------------------------------------

Note that Fig.~\ref{fig:ratio_xsec_lhc} implies also
that direct photon production is sensitive to the
large-$x$ PDFs, especially the gluon, but not to the small-$x$
ones: the inclusion of collider direct photon data into
a global PDF analysis might improve the precision of
the gluon at large-$x$, but not at small-$x$.

Let us finish the discussion on the impact
of high-energy resummation at the Tevatron
by noting that the origin of the discrepancy between NLO QCD
and experimental data at small $p_T$ is still not completely understood,
in particular, it is not caused by unresummed terms in the high-energy regime.
However, as we have discussed, since the direct photon cross section is much
more sensitive to large-$x$ effects, this discrepancy could
be partially cured by soft resummation~\cite{Becher:2009th}.

Now we turn to discuss the phenomenological impact of the resummation at LHC.
At the LHC, the production cross section of isolated photons is much
larger than at the Tevatron, which will make possible a  
high-statistics measurement. The ALICE, ATLAS, CMS and 
LHCb experiments at the LHC have photon reconstruction capabilities
 with the electromagnetic calorimetry  in various
 rapidity ranges~\cite{ddephotons}. 
The two main LHC experiments can measure photons in the central
rapidity region $|\eta^{\gamma}|\lsim 3$ down to $p_T=10$ GeV, ALICE
can do measurements  in the central
region $|\eta^{\gamma}|\lsim 0.7$ down to $p_T=5$ GeV, 
while LHCb can  measure forwards photons, $2\le |\eta^{\gamma}| \le 5 $ 
in the low $p_T\le$ 20 GeV region as well. The LHCb measurements are
specially interesting since small-$x$ resummation effects, which are only
important at low $p_T$, should be enhanced at forward rapidities.

\begin{figure}[t!]
\begin{center}
\epsfig{width=.8\textwidth,figure=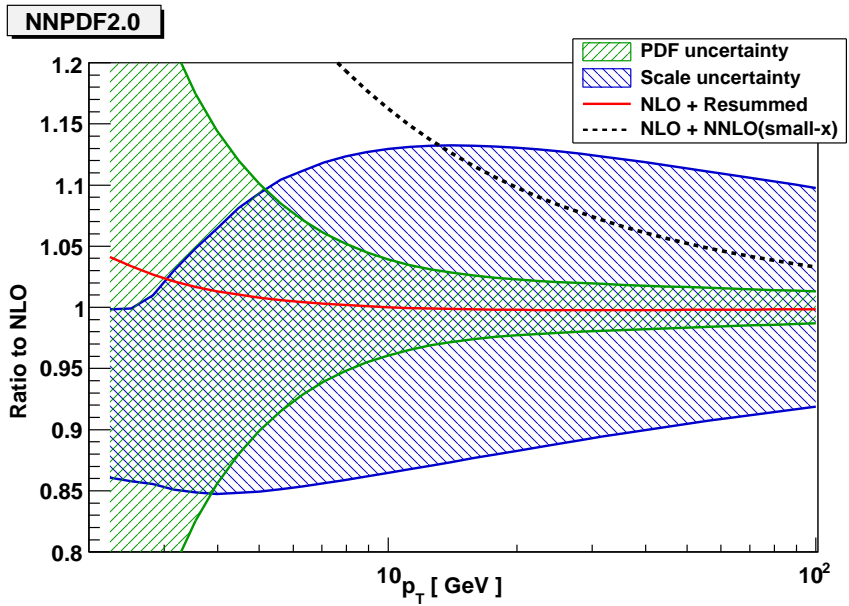} 
\caption{\small Ratio between resummed and NLO prediction
(solid red line) for the inclusive cross section at 
LHC, for a center of mass energy of $\sqrt{S}=14$ TeV. 
The NNPDF2.0 set has been
used to compute the theoretical prediction.
PDF and scale variation uncertainties are also shown. We
also show the ratio to NLO of the approximated NNLO result,
where the dominant NNLO contributions at small $x$ have been
added to the fixed order NLO result (black dashed line)
\label{fig:LHC}. 
}
\end{center}
\end{figure}

From the discussion in the case of
the Tevatron, we expect the impact of high--energy resummation
to be also small at the LHC.
To illustrate such impact,
in Fig.~\ref{fig:LHC} we show the ratio between the resummed and NLO 
direct photon production cross section at LHC, for
$\sqrt{S}=14$ TeV. We show for simplicity the direct
part of the photon production cross section only. No cuts
in the photon's rapidity are imposed. 
We have used again the NNPDF2.0
set for the theoretical prediction, and scale variation
uncertainty is estimated as discussed above. 

From Fig.~\ref{fig:LHC} we observe that the effect of high--energy
resummation is very small above $p_T\sim 10$ GeV, and it is
only for photon transverse momenta in the range 
$2~\GeV\lsim p_T \lsim 10~\GeV$
that it becomes of the order of a few percent. The origin
of the smallness of the high--energy resummation can be traced back,
as in the case of the Tevatron
to Fig.~\ref{fig:ratio_xsec_lhc}: the direct photon cross section for 
the production of a photon 
with $p_T=20$ GeV  is
completely dominated by the region $x\gsim 5\,10^{-3}$. In this
region, the resummed coefficient functions are almost
identical to the fixed order NLO ones. It is only for smaller
values of $p_T$ that the difference between NLO and resummed
coefficient functions at small-$x$, evident from
Fig.~\ref{fig:mycoef}, begin to contribute to the total cross section.
At very small-$p_T$ the effects of high--energy resummation
are much smaller than the PDF uncertainties.
This implies that the small-$p_T$ region can be used to constrain accurately
the gluon PDF, provided that systematic experimental uncertainties
in this region can be kept under control.

Let us emphasize however that the smallness of the high--energy resummation
with respect to fixed order NLO does not imply that resumming high--energy
enhanced terms is not relevant at hadronic
colliders. Indeed, the crucial role
of high--energy resummation is to  cure the instability 
of the cross section which appears in any fixed
order calculation at high--energy starting from NNLO. 
To illustrate this point, in Fig.~\ref{fig:LHC} we also
show the results for direct photon production if the dominant NNLO
contribution at small-$x$ (the term proportional to
$\mathcal{O}\lp \alpha_s^3\rp$ in Eq.~(\ref{eq:coeffres}) is
added to the fixed order NLO result, as an approximation
to the full fixed order NNLO result. We see that here 
the difference with respect NLO is more important, being
$\sim$10\% at $p_T\sim 20$ GeV and much larger at even
smaller $p_T$. The corresponding effect would be even larger
for the dominant NNNLO corrections. 
Thus the full high-energy resummation is required in
order to obtain stable predictions for future higher order
calculations of direct photon production (starting from NNLO accuracy) 
at small $p_T$ at hadronic colliders.

%------------------------------------------------------------
%------------------------------------------------------------------

%------------------------------------------------------------
\begin{figure}[t!]
\begin{center}
\epsfig{width=.8\textwidth,figure=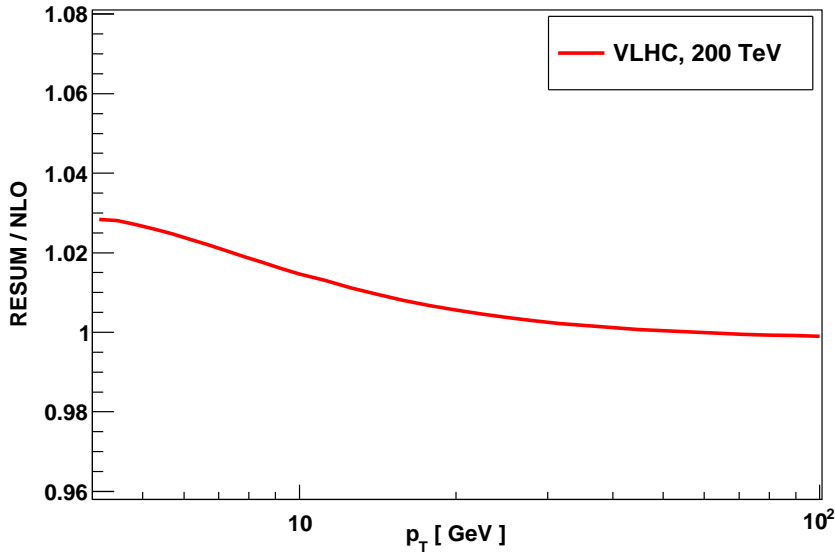} 
\caption{\small Ratio between resummed and NLO prediction
(solid red line) for the cross section for photon
production, integrated in rapidity, at 
a notional VLHC with a center of mass energy of $\sqrt{S}=200$ TeV. 
The NNPDF2.0 set has been
used to compute the theoretical prediction. Note that the very
large PDF and scale variation uncertainties are not shown
for simplicity.
\label{fig:VLHC} 
}
\end{center}
\end{figure}
%------------------------------------------------------------------

Finally, in Fig.~\ref{fig:VLHC} we show the impact of the resummation
of the high--energy coefficient function for photon production at a 
notional VLHC with $\sqrt{S}=200$ TeV. From Fig.~\ref{fig:ratio_xsec_lhc}
we see that for a 20 GeV photon the cross section
is sensitive to the coefficient functions
with $x \ge 5\,10^{-4}$, so one expected the effects
of the resummation to be more important that at
lower CM energies. However, even at this huge energy,
the effect is of a few percent at most at the
smallest $p_T$. 
%In this case PDF and scale variation
%uncertainties are not shown.
%Why not?

To summarize, in this letter results for the high
energy resummation of direct photon
production have been matched to NLO computations
and predictions for hadronic colliders have been
obtained. We have shown that main impact
of the full high--energy resummation procedure is to stabilize
 the logarithmic enhancement of
 the cross section at high energies which is present
at any fixed order in the perturbative expansion starting at NNLO. 
At the Tevatron the effects of the resummation
are completely negligible, while at the LHC high--energy resummation
of the partonic cross section enhances the hadronic cross section
be a few percent at small $p_T$, $p_T\lsim 10$ GeV. 
One important implication of our results is that 
the small $p_T$ discrepancy
between NLO QCD and Tevatron data cannot be described by unresummed
higher order contributions enhanced in the high--energy regime.
We have also shown that at the LHC the full resummation of the 
inclusive direct photo cross-section is very close
to the fixed order NLO QCD result, becoming significant only at 
very low $p_T$, and that even at a VLHC resummation effects are rather 
small in this channel.

%Finally, in this work we have discussed 
%the impact of the high--energy resummation
%of the coefficient function for the direct photon production. Our main
%result, that these corrections are very small at hadronic colliders,
%does not imply that high--energy resummation does not affect
%direct
%photon production as compared to the fixed order NLO
%result, since the effects which we have considered
%do not include the impact of
%PDF refitting. Indeed, contributions from PDFs
%obtained from a global small-$x$ resummed fit could still be 
%phenomenologically relevant. What our results imply is that
%when such small-$x$ resummed PDFs are available, a fully consistent
%high--energy resummation of the hadronic photon production cross
%section can be obtained by combining small-$x$ resummed PDFs with fixed order
%NLO partonic cross sections for all practical purposes.

\bigskip
\bigskip
\bigskip

{\bf \large Acknowledgements}\\
%\vspace{0.3cm}
We are especially grateful to S.~Forte for useful discussions and
continuous support during this project. We also thank 
D.~d'Enterria for illuminating 
discussions on photon production at hadronic
colliders, W.~Vogelsang and S.~Frixione for providing
us with their codes for photon production and for
assistance in using them, M.~Martinez for help
with the CDF data, and G.~Heinrich and J.~P.~Guillet for
discussions on photon production.
This work was partly supported 
by the European network HEPTOOLS under contract
MRTN-CT-2006-035505.
\bigskip

%\clearpage

%\bibliography{photons-pheno}

\input{photons-pheno.bbl}
\end{document}

%% file: preamble.tex
\usepackage{amsmath}
\usepackage{amssymb}
\usepackage{geometry}
\usepackage{epsfig}
\usepackage[usenames]{color}
\newcommand{\q}{\mathbf{q}}

\usepackage{soul}